# Inhomogeneous All-Dielectric Magnetic Metamaterials

**Authors:**  Jingbo Sun[1], Xiaoming Liu[2], Ji Zhou[2]*, Zhaxylyk Kudyshev[1], and Natalia M. Litchinitser[1]*

**Affiliations:**

[1]Department of Electrical Engineering, University at Buffalo, The State University of New York, Buffalo, New York 14260, USA

[2] *State Key Laboratory of New Ceramics and Fine Processing, School of Materials Science and Engineering, Tsinghua University, Beijing 100084, People's Republic of China*

*Correspondence to: natashal@buffalo.edu and zhouji@mail.tsinghua.edu.cn

**Abstract**: Anomalous field enhancement accompanied by resonant absorption phenomenon was originally discussed in the context of plasma physics and in applications related to radio-communications between the ground and spacecraft returning to Earth. Indeed, there is a critical period of time when all communications are lost due to the reflection/absorption of electromagnetic waves by the sheath of plasma created by a high speed vehicle re-entering the atmosphere. While detailed experimental studies of these phenomena in space are challenging, the emergence of electromagnetic metamaterials enables researchers exceptional flexibility to study them in the laboratory environment. Here, we experimentally demonstrated the localized field enhancement of the electromagnetic wave obliquely incident on a Mie-resonance-based inhomogeneous metamaterials with magnetic permeability gradually changing from positive to negative values.

**Main Text:** Radio frequency blackout phenomena are caused by a high electron concentration in the plasma sheath, generated around a high speed vehicle re-entering the atmosphere. As a result, there is a critical period of time when all communications are lost (*1-5*). An understanding of the interactions of electromagnetic waves with an ionized layer, including reflection, absorption, the effects of the gradient distribution of dielectric permittivity, and the width of the plasma sheath is of paramount importance for both fundamental science and mitigation of the blackout in practical radio-communication systems. While significant progress has been made in the development of theoretical models and numerical algorithms describing electromagnetic wave interaction with a given spatial density profile, experimental validations remain challenging. In this paper, we present the first experimental demonstration of anomalous field enhancement occurring near zero refractive index transition, using metamaterials. Moreover, we show that, in addition to anomalous field enhancement for the transverse-magnetic (TM) waves predicted in plasma physics, metamaterials provide an experimental platform for studying such phenomena for transverse-electric (TE) waves propagating in a medium with graded magnetic permeability.

The emergence of metamaterials provided researchers with exceptional flexibility for studying electromagnetic wave propagation in graded-index structures with respect to the spatial profile and absolute values of the refractive index distribution, as well as precise control over its distribution (*6-11*). Electromagnetic properties of metamaterials, such as their dielectric permittivities ($\varepsilon$), magnetic permeabilities ($\mu$), and refractive indices ($n$), can be designed to be positive, negative, or even zero at any selected frequency by properly adjusting the dimensions, periodicity, and other properties of their unit cells, or meta-atoms (*12-17*). Transition metamaterials (*9-11*)—artificial materials with the effective parameters, $\varepsilon$, $\mu$, or *n*, gradually changing from positive to negative values—have attracted significant and growing interest over the last few years. In particular, anomalous field enhancement and resonant absorption phenomena have been predicted for obliquely incident transverse-electric (TE) and/or transverse magnetic (TM) waves at the point where the real part of $\mu$ and/or $\varepsilon$ vanishes (*18-19*). These effects, besides most basic science aspects, hold the potential for numerous applications, including low-intensity nonlinear optics (*20, 21*), wave concentrators (*22-24*), and polarization-sensitive devices (25). Moreover, these phenomena may have important implications for different areas of surface and interface science (26).

In our recent theoretical studies (*27, 28*), we predicted a resonant field enhancement that occurs near the zero refractive index point under oblique incidence of the electromagnetic wave on an inhomogeneous metamaterial with magnetic permeability varying from positive to negative values, the so-called transition metamaterial. The physics of the phenomenon of resonant field enhancement in such graded-index metamaterials can be summarized as follows: for incident, transverse-magnetic, or transverse-electric waves, the thin layer near the zero-index point (transition point) can be considered a very thin capacitor or inductor that accumulates infinitely large electric or magnetic field energy, respectively, if we neglect the effects of dissipation and spatial dispersion. Note that this energy accumulation takes place only in the case of obliquely incident waves, as only in this case the electric (magnetic) field at the oblique incidence has a non-zero component in the direction of propagation. Since electric displacement D (magnetic induction B) must be continuous, the electric field E (magnetic field H) anomalously increases as $\varepsilon$ ($\mu$) tends to zero.

In this study, we designed and fabricated graded magnetic permeability metamaterials using a high-index ceramic cube array with graded lattice constants such that the magnetic permeability gradually changes from a positive to a negative value along the propagation direction (Fig. 1A). As opposed to previously realized graded-index structures, such as electromagnetic cloaks that were based on resonant metallic resonators embedded in a dielectric matrix, we designed all-dielectric transition metamaterials. While this first experimental demonstration reported here was performed at microwave frequencies, where metal losses are negligible, an all-dielectric design will be advantageous at optical frequencies, where metallic losses are significant and, moreover, plasmonic enhancement at metal edges may mask the effects of the anomalous field enhancement discussed above.

**Theory**. The enhancement effect in the transition metamaterial can be understood by solving the wave equations inside the transition metamaterial. Assume that a TE $(E = \{0, 0, E_z\}, H = \{H_x, H_y, 0\})$ wave at an incident angle of $\theta$ is propagating from air (*x*<0) into a transition medium with a width of $2h$ $(-h < x < h)$, where relative dielectric permittivity $\varepsilon$ is a constant $\varepsilon_{r0}$, and relative magnetic permeability $\mu$ has a graded profile given by $\mu(x) = \mu_{r0} f(\xi)$, where $\xi = x/h$ is changing from positive to negative, with zero point at $\xi = 0$, $f(\xi = 0) = 0$. Since the medium is homogeneous in the *y* direction, the electric field component can be written as $E_z = G(x)\exp(i\omega t - ik_y y)$, where $k_y = (\varepsilon_{r0} \mu_{r0})^{1/2} (\omega/c)\sin(\theta)$. Then, using the system of Maxwell

equations, we can write the Helmholtz equation for such an inhomogeneous medium in the following form:

$$\frac{d^2G}{d\xi^2} - \frac{1}{\mu(\xi)}\frac{d\mu(\xi)}{d\xi}\frac{dG(\xi)}{d\xi} + \frac{\omega^2 \varepsilon_{r0}\mu_{r0}h^2}{c^2}\left(\xi - \sin^2(\theta)\right)G(x) = 0 \quad (1)$$

Where $\mu(\xi) = -\mu_{r0}\xi$, $E_z$ is the amplitude of the electric component of the harmonic EM wave at frequency $\omega$, and $c$ is the speed of light in the vacuum. The magnetic field component along the propagation direction, $H_x$ can be calculated, using Maxwell's equations, as

$$H_x = -\frac{ic}{\omega\mu(\xi)}\frac{\partial E_z}{\partial y}. \quad (2)$$

Noticing that $E_z$ depends on the coordinate $y$ as $E_z = G(x)\exp(i\omega t - ik_y y)$ and assuming $\mu(\xi) = -\mu_{r0}\xi$ (neglecting material losses), we obtain $H_x = \frac{ck_y}{\omega\mu_{r0}\xi}E_z$.

While in the normal incidence case, $\theta = 0$, no unusual behavior is predicted for both electric and magnetic field components, in the case of oblique angle ($\theta \neq 0$), a strong enhancement for the longitudinal component of $H_x$ at the position where permeability is changing from positive to negative is predicted. Figure 1B shows the results of numerical simulations, confirming this prediction.

**Experiment**. In order to demonstrate the predicted field enhancement in the vicinity of the zero-$\mu$ transition point, we designed and fabricated all-dielectric metamaterials made of high-refractive-index dielectric cubes. Such cubes have been shown to produce electromagnetically induced electric and magnetic resonances. A magnetic resonance originates from the excitation of a particular electromagnetic mode inside the cube with a circular displacement current of the electric field. In contrast to more conventional, split-ring resonator based metamaterials that possess anisotropic electromagnetic response (*6-8*), the properties of the all-dielectric structures demonstrated here are isotropic.

The transition metamaterial was realized using a high-refractive-index cube array with graded lattice constants (*17*), so that it possessed a graded magnetic permeability profile, varying from positive to negative values, as shown in Fig. 2. The high-index cubes with a side length of 2.2mm were made from a dielectric ceramic material, $CaTiO_3$, doped with 15 wt % $ZrO_2$, which was synthesized by the solid state reaction method under 1400 °C, and the measured permittivity of it was 122+0.244i. The lattice constant was gradually changing from 6 mm to 5 mm over the length of 66 mm.

Using commercial, full-wave, finite-element simulation software (Microwave Studio, Computer Simulation Technology), we performed a series of scattering (S) parameter simulations for the dielectric cube, with different unit cells' sizes $c$ covering the X band as shown in Fig. 2A. The standard retrieval procedure (*29, 30*) was used to obtain the effective material parameters, as shown in Fig. 2B. In our design, 10.472 GHz was chosen as the operating frequency. Then, the array was designed such that the effective magnetic permeability of an array would change from 0.16 to -0.31 along the *x* direction.

The transition metamaterial sample comprises an array consisting of 12 rows of ceramic cubes. Every two consecutive rows possess the same unit cell sizes, varying in range from 6 mm to 5 mm, as shown in Fig. 2C. Therefore, from bottom to top, permeability varies gradually from positive to negative in the frequency range around the operating frequency. In the experiments, these cubes were imbedded in an ABS matrix, which makes the entire structure firm and steady.

We utilized the 2D near-field scanning system to measure the power distribution. The corresponding H-field, for which the enhancement was theoretically predicted, was then calculated based on the measurement results. As shown in Fig. 3A, the near-field scanning system consists of two parallel metal plates with a gap of 11 mm between them, forming a plane waveguide that supports the mode, with an E field pointing in the *z* direction in the X band. Absorbers on the lower metal plate restrict the measurement area and the path for feeding the incident wave. Microwaves spanning across the X band (8–12GHz) that includes the operating frequency of the designed sample are generated by the vector network analyzer (VNR), Agilent ENA 5071C. A detecting probe, which is also connected to the VNR, is installed in the center of the upper metal plate to collect the amplitude and phase of the E field. The transition metamaterial sample was placed in the middle of the lower plate with an angle of 30 degrees between its normal and the incident direction. During the measurement, the sample was moved with the bottom plate along the U and V directions with a scanning step of 3 mm, and a full, 2D spatial field map of the microwave scattering pattern was acquired, both inside the transition metamaterial and in the surrounding free-space region.

The measurement results of the power, shown in Fig. 3B, can be used to calculate the field $E_z$ and then transformed into the $H$ field, according to Eq. 2. The calculated results of the longitudinal field component ($H_x$) perpendicular to the boundary of the metamaterial are shown in Fig. 3C. At the region around the arrays with zero permeability, $H_x$ is greatly enhanced, which agrees well with the prediction in Fig. 1.

Therefore, we experimentally observed a high enhancement of the H field for an electromagnetic wave propagating in inhomogeneous metamaterials with permeability linearly changing from positive to negative values. Although these preliminary results were obtained in the microwave frequency range, the proposed all-dielectric approach to transition metamaterials can be extended to infrared and optical frequencies. An important advantage of the all-dielectric approach in this spectral range is that unlike plasmonic metal-dielectric meta-atoms, all-dielectric cubes do not suffer conduction losses at optical frequencies.

**Acknowledgments:** We thank Qian Zhao and Zongqi Xiao for the help with the near field scanning measurement. This research was supported by the US Army Research Office Awards W911NF-11-1-0333, W911NF-15-1-0146 and National Natural Science Foundation of China 11274198, Beijing Municipal Natural Science Program Z141100004214001.

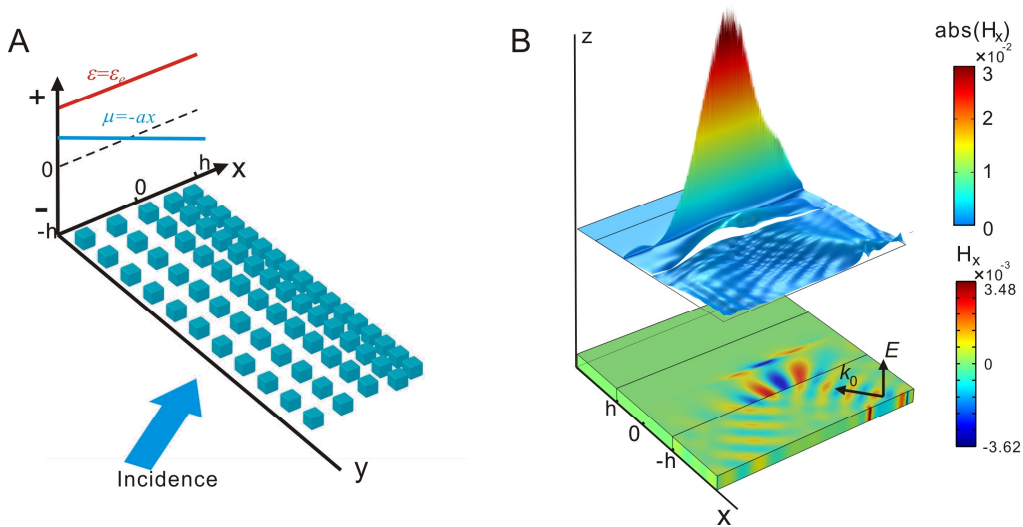

**Fig. 1**. Field enhancement in a transition metamaterials: (A) Schematic of the transition layer; (B) Field enhancement effect in transition metamaterials with varying magnetic permeability.

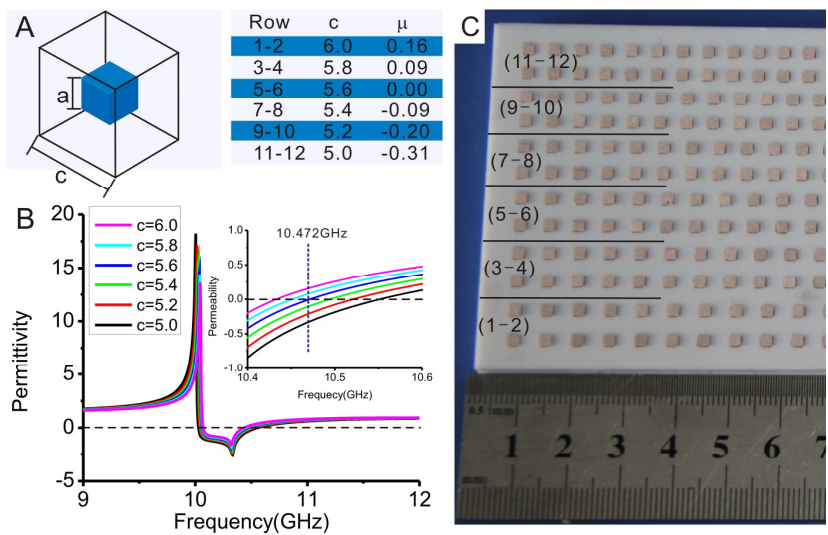

**Fig. 2.** Effective parameters of the transition metamaterials: (A) Schematic illustration of the unit cell; (B) Retrieved parameters of the transition metamaterials made of identical ceramic cubes with side length of $a$=2.2mm and varying unit cell size $c$; (C) The experimental sample of transition material, comprised of ceramic cubes with graded lattice constants.

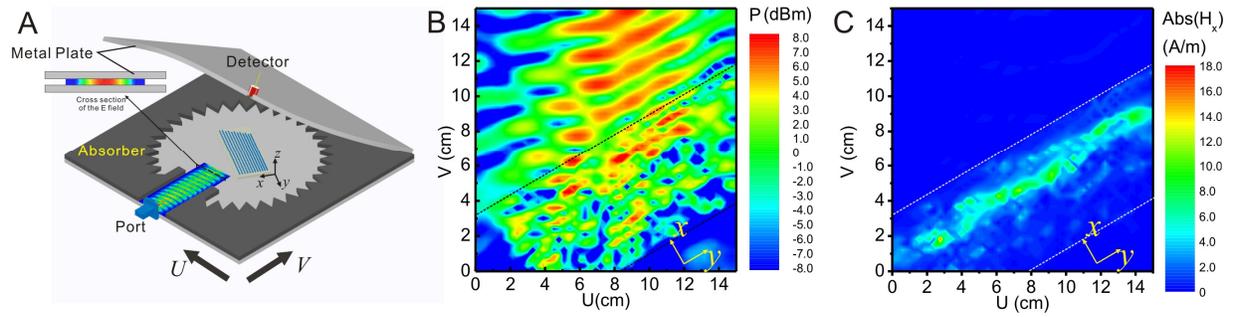

**Fig. 3.** Measurement setup and results: (A) The schematic of the near-field scanning system; (B) The measured power distribution at 10.415 GHz under oblique incidence of the electromagnetic wave on the transition metamaterials; (C) The distribution of the $H_x$ field component, calculated from the experimentally measured power distribution. The dash lines in (B) and (C) show the boundaries of the sample.